\documentstyle[prl,preprint,aps,epsf]{revtex}

\begin{document}


\title{The instability of Alexander-McTague crystals and its
  implication for nucleation}

\author{W.\ Klein} 

\address{Center for Non-Linear Studies, Los Alamos National Laboratory, Los Alamos,
NM 87545}

\address{Department of Physics and Center for
Computational Science${}^{\dag}$\\ Boston University, Boston, MA 02215}

\maketitle

\begin{abstract} We show that the argument of Alexander and McTague,
  that the bcc crystalline structure is favored in those
  crystallization processes where the first order character is not too
  pronounced, is not correct. We find that any solution that satisfies
  the Alexander-McTague condition is not stable. We investigate the
  implication of this result for nucleation near the pseudo-spinodal
  in near-meanfield systems. 
\end{abstract}

\noindent
${}^{\dag}$ Permanent address

\section{Introduction}
Predicting the symmetry of crystalline phases from first principles
and the knowledge of the inter-molecular potential has eluded
condensed matter physicists for years. One promising path was
suggested by Alexander and McTague(AM)\cite{am} who noted that the
Landau-Ginsburg free energy indicated a particular crystalline form if
the first order nature of the transition from the liquid was not too
pronounced. For systems such as those that interact with a
Lennard-Jones(LJ) potential the AM argument indicated that the
crystalline phase would have a triangular or hexagonal structure 
in $d=2$ while the same argument suggested that the bcc lattice would
be favored in $d=3$. This conclusion is independent of the
details of the potential as long as, like LJ, it is spherically 
symmetric and the fluid is monatomic.
However, simulations of LJ and hard sphere systems as well as
experiments on metals and rare gasses indicate that these systems
freeze into an fcc structure.\cite{ten,refs} This result is also obtained by 
the application of density functional theories.\cite{sin,lo}   

In a related development
Klein and Leyvraz(KL)\cite{KL} used the AM argument to look at
nucleation of the crystal from the meta-stable liquid in a mean-field
system near the spinodal and found a hexagonal or triangular  
structure for the critical droplet in $d =2$ and stacked triangular or
hexagonal planes or a bcc structure in $d = 3$. 
However, this droplet was unstable. In addition, Groh and
Mulder(GM)\cite{GM} also
looked at a mean-field system near a spinodal and, using bifurcation
analysis, also noted that the AM argument said nothing about the
nature of the stable solid but only indicated that a bifurcation to a
bcc structure in d=3 was favored over bifurcations to other periodic 
structures but the stability of the bcc
structure was beyond the AM or bifurcation analysis. From these
results it is still unclear what the meaning or accuracy is of the AM
argument. It is the purpose of this paper to perform some additional
analysis of these results.  

In this paper we present an argument indicating that any crystal
satisfying the conditions in AM must be unstable. This argument, 
which uses the Landau-Ginsburg approach employed by AM does not
require that the system be mean-field. In addition we use the
techniques of AM to investigate the nucleation process near the
spinodal like singularity in systems with long range interactions. We
argue that the droplet with the structure indicated by the AM
argument dominates the nucleation process. The fact that this droplet
is unstable is, unlike the phases proposed by AM, physically
reasonable and consistent with other results. In addition we
investigate the bifurcation analysis of GM and show that the
additional bifurcations to other symmetries that they find, which are
not found by AM are not relevant to nucleation.

The remainder of this paper is
structured as follows; In Section II we describe the method of AM. In
Section III we present the argument for instability. In Sections IV 
and V we
discuss the implcations of this argument for nucleation of the 
crystal from the liquid and in Section VI we summarize and discuss  
our results.

\section{Alexander-McTague}

In this section we give a brief description of the argument of AM.\cite{am} 
We will use a slightly different, but equivalent approach. 
We begin with the Landau-Ginsburg free energy
\begin{equation}
\label{lg}
F(\rho)= {1\over 2}\int\int C(|{\vec x} - {\vec y}|)\rho({\vec x})\rho({\vec
  y})d{\vec x}d{\vec y} + \int f(\rho({\vec x}),T)d{\vec x} -
  h\int\rho({\vec x})d{\vec x}
\end{equation}
In the above we assume, following AM, that the interaction term
is quadratic in $\rho$ and that $f(\rho,T)$ can be expanded in a power
series in $\rho$ for fixed $T$. The quantity 
$h$ is the chemical potential. The Euler-Lagrange
equation that specifies the equilibrium state is obtained by
functionally differentiating eq.(\ref{lg}) and setting the derivative
equal to zero. 
\begin{equation}
\label{euler}
\int C(|{\vec x} - {\vec y}|)\rho({\vec y})d{\vec y} + {{\partial
    f(\rho,T})\over {\partial \rho}} - h = 0
\end{equation}
We assume that there is a constant solution $\rho_{o}$ corresponding to
the stable or metastable liquid at the values of $T$ and $h$ we are
investigating. Defining $\psi({\vec x})$ through $\rho({\vec
  x})=\rho_{o}+\psi({\vec x})$ and inserting this expression in
eq.(\ref{euler}) we obtain
\begin{equation}
\label{eulers}
\int A(|{\vec x} - {\vec y}|)\psi({\vec y})d{\vec y} +
b(h,T)\psi^{2}({\vec x}) = 0
\end{equation}
Here $b(h,T)$ is the coefficient of the term quadratic in $\psi({\vec
  x})$ arising out of a Taylor series  expansion of 
${{\partial f(\rho)}\over {\partial \rho}}$. The
  dependence of $b(h,T)$ on $h$ and $T$ is through $\rho_{o}$. The term
$A(|{\vec x} - {\vec y}|) = C(|{\vec x} - {\vec y}|) +
  b_{1}\delta({\vec x} - {\vec y})$ where $b_{1}$ is the coefficient of
  the term linear in $\psi({\vec x})$ in the Taylor
  expansion which also depends on $h$ and $T$ through $\rho_{o}$ and
  $\delta({\vec x} - {\vec y})$ is the Dirac delta function. 
Following AM we have truncated the Taylor expansion after
the quadratic term using the assumption that $\psi({\vec x})<<1$,
i.e. the first order nature of the transition is not too pronounced. 
We will see that this can be justified in a self consistent manner.            

Again following AM we consider the system in the neighborhood of a
critical point associated with the fluid.\cite{am} The critical point
is characterized by ${\hat A}(|{\vec k}_{o})|)=0$, where $|{\vec k}_{o}|$ is 
the location of the global minimum of ${\hat C}(|({\vec k}|)$ defined in
eq.(\ref{lg}). Near the critical point ${\hat A}(|{\vec k}_{o}|)=\epsilon<<1$.
The hat denotes the Fourier Transform and $\epsilon = {{T-T_{S}}\over
  T_{S}}$ where $T_{S}$ is the critical temperature. 
From eq.(\ref{eulers}) we obtain, after Fourier transform,
\begin{equation} 
\label{eulersft}
{\hat A}(|{\vec k}|){\hat \psi}({\vec k}) + b(h,T)\int {\hat
  \psi}({\vec k}-{\vec k}^{\prime}){\hat \psi}({\vec k}^{\prime})d{\vec
  k}^{\prime} = 0 
\end{equation}
There is one solution of eq.(\ref{eulersft}) that scales with
$\epsilon$ i.e. ${\hat \psi}({\vec k})\sim \epsilon$ or
$\psi({\vec x})\sim \epsilon$. The scaling
justifies neglecting the higher order terms in $\psi$ since they will
be higher order in $\epsilon$. Such scaling solutions, assuming of
course that they exist, will have a free energy cost which differs
from the liquid by terms of order $\epsilon$. What about solutions 
that do not scale? Unless the critical
point $T_{S}$ is on the coexistence curve it seems likely that there
would be solutions with a lower free energy than the liquid as 
$\epsilon\rightarrow 0$. This existence of other solutions
is what one would expect if, as we contend, the AM solution is
unstable.  
However, here we are simply following the AM argument and do
not need to consider the possibility of solutions of the
Euler-Lagrange equation that do not scale and hence require higher
order terms in the Landau-Ginsburg free energy then those allowed by AM.   
The question of solutions for nucleation droplets that do
not scale with $\epsilon$ will be addressed in Section V. 

We now assume that the solution of eq.(\ref{eulers}) is of the form
\begin{equation}
\label{soln}
\psi({\vec x}) = \sum_{n}c_{n}(h,T)\exp(i{\vec k}_{n}\cdot{\vec x})
\end{equation}
where the ${\vec k}_{n}$ are reciprocal lattice vectors. If we now
take $d$ reciprocal lattice vectors, e.g. ${\vec k}_{1}... {\vec
  k}_{d}$ where $d$ is the spatial dimension, to have a magnitude
$|{\vec k}_{o}|$ then we have a solution in which $c_{1}... c_{d}>> c_{n}$ 
for $n>d$ when $\epsilon<<1$; that is for values of $h$ and $T$ near
the critical point.

To see this we simply insert eq.(\ref{soln}) into eq.(\ref{eulers}) to
obtain 
\begin{equation}
\label{soln2}
\sum_{n=1}^{\infty} c_{n}(h,T){\hat A}(|{\vec k}_{n}|)\exp(i{\vec
  k}_{n}\cdot {\vec x}) + b(h,T)\bigl\lbrack\sum_{n=1}^{\infty}
  c_{n}(h,T)\exp(i{\vec k}_{n}\cdot {\vec x})\bigr\rbrack^{2} = 0
\end{equation}
We now take $c_{1}(h,T)....c_{d}(h,T) \sim \epsilon$ and all other 
$c_{n}(h,T)$ to be higher order in $\epsilon$. We can assume ${\hat
  A}(|{\vec k}_{n}|)\sim 1$ for $n \neq 1,..., d$. It is straightforward
to generalize this argument to the case of a finite number of
degenerate global minima. For $\epsilon<<1$ the modes with magnitude
$|{\vec k}_{o}|$ dominate the solution. In the limit
$\epsilon\rightarrow 0$ these modes are the only ones that contribute  
Since these modes dominate for $\epsilon<<1$ they must be such that 
$(\sum_{n=1}^{d}\exp(i{\vec k}_{o,n}\cdot {\vec x}))^{2}$ 
has the same symmetry as $\exp(i{\vec k}_{o,n}\cdot {\vec x})$.
Here ${\vec k}_{o,n}$ is one of the $d$ reciprocal lattice vectors
with magnitude $|{\vec k}_{o}|$. Note that the sum contains
exponentials and their complex conjugates as the sum must be real. 
Therefore as in AM we must have that
the reciprocal lattice vectors form equilateral trangles. In two
dimensions this generates a triangular or hexagonal lattice in real 
space and in three dimensions either bcc or layered triangular or
hexagonal structures. This result was also derived via a bifurcation
analysis by GM.\cite{GM} In the next section we investigate the
stability of the solutions of eq.(\ref{eulers}).

\section{Stability of the Alexander-McTague Solutions} 
To check the stabilty of the solutions of eq.(\ref{eulers})
we perform a simple linear stability analysis. Writing $\psi({\vec
  x})+\eta({\vec x})$, where $\eta({\vec x})$ is a small arbitrary 
perturbation and $\psi({\vec x})$ is a solution to eq.(\ref{eulers}), 
and inserting this in eq.(\ref{eulers}) we obtain
\begin{equation}
\label{inst1}
\int A(|{\vec x} - {\vec y}|)\eta({\vec y})d{\vec y} +
2b(h,T)\psi({\vec x})\eta({\vec x}) = \Phi(\eta({\vec x}))
\end{equation}
The function $\Phi(\eta({\vec x}))$ can be expanded in the 
eigenvectors of the operator on the left hand side of eq.(\ref{inst1}).
If the eigenvalues of the operator are all positive then the solution
is linearly stable. If there is at least one negative eigenvalue 
the solution is linearly unstable. We will show that there must exist
at least one negative eigenvalue. 

The proof makes use of the well known fact that if we have an
Hermetian operator $\Theta$ defined on a Hilbert space then 
the average of $\Theta$, which we will call ${\bar
  \Theta}$, which is defined as 
\begin{equation}
\label{bound1}
{\bar \Theta} = {<x|\Theta|x>\over <x|x>}
\end{equation}
is an upper bound for the lowest eigenvalue of the operator $\Theta$. Here
$|x>$ is any vector in the Hilbert space. The derivation of this
result is quite straightforward and can be found in {\it Modern
  Quantum Mechanics} by J. J. Sakurai\cite{sak} as well as most
elementary books on Quantum Mechanics. Note that the operator in
eq.(\ref{inst1}) is both real and symmetric and hence Hermetian. The
Hilbert space we use is defined with plane wave normalization. 

Since we are primarily interested in the situation in which the fluid
phase with density $\rho_{o}$ is stable or metastable we begin with
this case first. Since ${\hat A}({\vec k})$ is the ${\vec k}$
dependent susceptibility of the uniform fluid if the fluid is stable
or metastable then 
${\hat A}({\vec k})>0$ for all $|{\vec k}|$. Using 
$\psi({\vec x})$, the solution to eq.(\ref{eulers}), as the vector
$|x>$ in our bound we have that the bound $B$ is given by
\begin{equation}   
\label{bound2}
B = {\int \psi^{\star}({\vec x})A(|{\vec x} - {\vec y}|)\psi({\vec
    y})d{\vec x}d{\vec y} + 2b(h,T)\int \psi^{\star}({\vec
    x})\psi^{2}({\vec x})d{\vec x}\over \int |\psi({\vec
    x})|^{2}d{\vec x}}
\end{equation}
Since $\psi({\vec x})$ is a solution to eq.(\ref{eulers}) the bound
$B$ reduces to
\begin{equation}
\label{bound3}
B = {-\int \psi^{\star}({\vec x})A(|{\vec x} - {\vec y}|)\psi({\vec
    y})d{\vec x}d{\vec y}\over \int |\psi({\vec x})|^{2}d{\vec x}}
\end{equation}
Converting to Fourier space by using Parseval's theorem we have
\begin{equation}
\label{bound4}
-B\int |{\hat \psi}({\vec k})|^{2}d{\vec k} = \int |{\hat \psi}({\vec
 k})|^{2}{\hat A}(|{\vec k}|)d{\vec k}
\end{equation}
Since ${\hat A}(|{\vec k}|)$ is assumed to be positive definite we
must have $B<0$. Hence, the upper bound of the lowest eigenvalue is
less than zero which proves the result. Note the fact that
$\psi({\vec x})$ is periodic was never used in the argument which is
therefore valid for any solution of eq.(\ref{eulers}).

Although it is not relevant for the AM argument, where the question of
the structure of the solid phase near a stable or metastable liquid
was considered,
for the sake of completeness we demonstrate that the solutions of
eq.(\ref{eulers}) are unstable when ${\hat A}(|{\vec k}|)<0$
for some value(s) of $|{\vec k}|$; that is, when the liquid phase is
unstable. Equation \ref{bound1} for the lowest eigenvalue bound
can be written as
\begin{equation}
\label{bound5}
B\int |w({\vec x})|^{2}d{\vec x} = \int w^{\star}({\vec x})A(|{\vec x}
- {\vec y}|)w({\vec y})d{\vec x}d{\vec y} + 2b(h,T)\int \psi({\vec
  x})w^{\star}({\vec x})w({\vec x})d{\vec x}
\end{equation}
where the operator $\Theta$ is given in eq.(\ref{inst1}). We choose the
test funtion $w({\vec x}) = \exp(i{\vec k}_{o}\cdot {\vec x})$ so that 
\begin{equation}
\label{bound6}
BV = {\hat A}(|{\vec k}_{o}|)V + 2b(h,T)C
\end{equation}
where $C = \int \psi({\vec x})d{\vec x}$ and $V$ is the system volume.
Since ${\vec k}_{o}$ is the location of the global minimum of ${\hat
  A}(|{\vec k}|)$ and, by assumption, ${\hat A}(|{\vec k}|)<0$ for
some $|{\vec k}|$, ${\hat A}(|{\vec k}_{o}|)<0$. Returning to
eq.(\ref{eulers}) multiplying by $b(h,T)$ and integrating with respect
to ${\vec x}$ yields
\begin{equation}
\label{bound7}
b(h,T)C = {-b^{2}(h,T)\int|\psi^{2}({\vec x})|d{\vec x}\over {\hat
    A}(0)}
\end{equation}
Substituting eq.(\ref{bound7}) into eq.(\ref{bound6}) we obtain
\begin{equation}
\label{bound8}
BV = {\hat A}(|{\vec k}_{o}|)V - {2b^{2}(h,T)\int |\psi({\vec
    x})|^{2}d{\vec x}\over {\hat A}(0)}
\end{equation}
Since $b(h,T)$ is real and ${\hat A}(|{\vec k}_{o}|)$ is assumed 
to be negative, if ${\hat   A}(0)>0$ then $B < 0$. If ${\hat A}(0) < 0$
eq.(\ref{eulers}) can no
longer be used to generate an equlibrium crystal. This occurs because 
$A({\vec x} - {\vec y})$ and $b(h,T)$ are functions of $\rho_{o}$
which is unstable to spatially constant perturbations.

The one situation left to address is the one in which ${\hat A}(|{\vec
  k}|)$ is positive for all $|{\vec k}|$ except for $|{\vec k}_{o}|$
  where it is equal to zero. This is the situation in which the system
  is at the critical point. From eq.(\ref{bound4}) we now have $B=0$
  if ${\hat \psi}({\vec k}) = \delta({\vec k} - {\vec k}_{o})$. This
  implies that $\psi({\vec x}) = \exp(i{\vec k}_{o}\cdot{\vec
  x})$. However, from our discussion in Section II $\psi({\vec
  x})\sim\epsilon = {\hat A}(|{\vec k}_{o}|) = 0$. Therefore the
  solution of eq.(\ref{euler}) in this case describes the infinitely
  long lived fluctuations at the critical point and not a phase. 

Note that the
  condition that we be near a critical point (${\hat A}(|{\vec k}|) =
  \epsilon<<1$ for some $|{\vec k}|$) was necessary to argue that the
  AM crystals were in fact minima of the free energy. However, the
  critical point condition was not used to show that the solution of
  eq.(\ref{eulers}) was unstable. In addition, we have not specified the nature
  of the critical point. When the critical point is a spinodal then
  the AM argument becomes quite useful in determining the nature of
  the critical droplet in the nucleation process that takes place near
  the pseudo-spinodal in near-meanfield systems. We discuss this
  application in the next section.

Note also that the same argument that was used to show that the
solutions of eq.(\ref{eulers}) are unstable can be used on any
equation of the form
\begin{equation}
\label{eulergen}
\int A(|{\vec x} - {\vec y}|)\psi({\vec y})d{\vec y} +
q(h,T)\psi^{n}({\vec x}) = 0 
\end{equation}
where $n$ in any integer greater than or equal to two.
 
\section{Spinodal Nucleation of a Crystal from the Liquid}
In this section we apply the ideas of AM to the problem of the
nucleation of a crystalline solid from the liquid near a 
pseudo-spinodal in a near-meanfield system. First we note that
spinodals, in the sense that one finds them in the van der Waals
theory of liquids or the Curie-Weiss theory of magnetic systems, are
meanfield(MF) objects. In Ising models this has been seen in Monte
Carlo simulations\cite{heerm} and via transfer matrix
techniques.\cite{novot} A general argument using a Ginsburg criterion  
has also
been given by Binder\cite{bin}. To study MF systems in a rigorous
way Kac\cite{kac} introduced the idea that, if a system has an
interaction potential of the form $V(|{\vec x}|) = V_{ref}(|{\vec x}|)
  + \gamma^{d}\phi(\gamma |{\vec r}|)$ where $V_{ref}(|{\vec x})$ is a
  short range reference potential, $\gamma$ is a parameter and $\int
  \gamma^{d}\phi(\gamma|{\vec x}|)d{\vec x} = D<\infty$, then in the
  limit $\gamma\rightarrow 0$ the system will be MF. This means that
  for fluids such a potential results in the van der Waals equation
  with the attendant MF critical exponents and spinodals. In magnetic
  systems the result is the Curie-Weiss description. We have assumed
  that the interaction is spherically symmetric for simplicity but
  that need not be the case.

In order to generate near-meanfield(NMF) systems we use the approach of Kac
however we take $\gamma<<1$ but finite. That is we do not take the 
$\gamma\rightarrow 0$ limit. In NMF systems there is no true spinodal
but the, depending on the interaction range $R=\gamma^{-1}$ the system
will behave as if there is a spinodal as long as one does not approach
the singularity too closely.\cite{heerm,novot,bin} We will refer to
such apparent singularities as pseudo-spinodals. 
To study nucleation in systems with long range interactions undergoing 
deep quenches we adopt the techniques of saddle point evaluation
of the partition function\cite{lang} to nucleation near the pseudo-spinodal.
\cite{KL,UK} We begin with a Landau-Ginsburg-Wilson Hamiltonian
identical to $F(\rho)$ in eq.(\ref{lg}) but with one additional
requirement. We take the interaction term
\begin{equation}
\label{int}
C(|{\vec x} - {\vec y}|) = \gamma^{d}\Lambda(\gamma|{\vec x} - {\vec y}|)
\end{equation}
where the $\Lambda$ has the properties of the long range Kac potential
described above. We will take $\gamma$ to be small but finite so that
we are describing NMF rather than MF systems. 
Our Hamiltonian $H$ is then
\begin{equation}
\label{ham1}
H(\rho) = {1\over 2}\int \gamma^{d}\Lambda(\gamma|{\vec x} - {\vec
  y}|)\rho({\vec x})\rho({\vec y})d{\vec x}d{\vec y} +
  \sum_{n=1}^{\infty}\int b_{n}(h,T)\rho^{n}({\vec x})d{\vec x} -
  h\int \rho({\vec x})d{\vec x}
\end{equation}
where we have made the Taylor series expansion of $f(\rho,T)$
explicit. The partition function in the canonical ensemble is
\begin{equation}
\label{part}
Z = \int \delta \rho \exp(-\beta H(\rho))
\end{equation}
where $\beta = 1/K_{B}T$.

Using $R = \gamma^{-1}$ and assuming that $\rho({\vec r}) = \rho({\vec
  x}/R)$ the Hamiltonian $H$ in terms of scaled lengths becomes
\begin{equation}
\label{ham2}
H(\rho) = R^{d}\biggl\lbrack {1\over 2}\int \Lambda(|{\vec r} - {\vec
  r}^{\prime}|)\rho({\vec r})\rho({\vec r}^{\prime})d{\vec r}d{\vec
  r}^{\prime} +\sum_{n=1}^{\infty} b_{n}(h,T)\int \rho^{n}({\vec
  r})d{\vec r} - h\int \rho({\vec r})d{\vec r}\biggr\rbrack
\end{equation}
For $R>>1$ (i.e. the NMF limit) the partition function in
eq.(\ref{part}) can be evaluated as a saddle point integral. The
Euler-Lagrange equation that specifies the saddle point will be of the
same form as eq.(\ref{euler}). Identifying $\rho_{o}$ as the constant
solution that specifies the density of the liquid we write $\rho({\vec
  r}) = \rho_{o} + {\bar \psi}({\vec r})$. As in AM we assume that
${\bar \psi}({\vec r})$ is small since we are near a pseudo-spinodal
critical point. Expanding in ${\bar \psi}({\vec r})$ we obtain an
equation for ${\bar \psi}({\vec r})$ of the same form as eq.(\ref{eulers}). 
\begin{equation} 
\label{eulern}
R^{d}\bigl\lbrack\int {\bar A}(|{\vec r} - {\vec r}^{\prime}|){\bar
  \psi}({\vec r}^{\prime})d{\vec r}^{\prime} + b(h,T){\bar
  \psi}^{2}({\vec r})\bigr\rbrack = 0
\end{equation}
where ${\bar A}(|{\vec r} - {\vec r}^{\prime}|) = \Lambda(|{\vec r} -
{\vec r}^{\prime}|) + b_{1}\delta({\vec r} - {\vec r}^{\prime})$ 

To obtain the nucleation droplet we assume a solution of the form 
\begin{equation}
\label{drop}
{\bar \psi}({\vec r}) = 
\sum_{n} \exp(i{\vec k}^{\prime}_{o,n}\cdot{\vec r}){\bar \psi}^{\prime}({{\vec
      r}\over L})
\end{equation} 
where ${\vec k}_{o}^{\prime} = R{\vec k}_{o}$  
and $L$ is a length to be determined. We assume at the outset that
$L>>|{\vec k}_{o}|^{-1}$, which will be seen to be true self consistently. 
As before $|{\vec k}_{o}|$ is the location of the global minimum of
${\hat{\bar A}}(|{\vec k}|)$. Since $|{\vec k}_{o}|\sim R$, ${\bar
  \psi}^{\prime}({{\vec r}\over L})$ is a slowly varying function of
${\vec r}$. Consequently, we can expand ${\bar \psi}^{\prime}({{\vec
    r}^{\prime}\over L})$ in a gradient expansion about ${\vec
  r}^{\prime} = {\vec r}$. The first three terms in the expansion are
\begin{equation}
\label{expan}
{\bar \psi}^{\prime}({{\vec r}^{\prime}\over L}) = {\bar
  \psi}^{\prime}({{\vec r}^{\prime}\over L}) + {({\vec r} - {\vec
  r}^{\prime})\over L}\cdot \nabla {\bar \psi}^{\prime}({{\vec
  r}^{\prime}\over L})|_{{\vec r}^{\prime}={\vec r}} + {|{\vec r} -
  {\vec r}^{\prime}|^{2}\over L^{2}}\nabla^{2}{\bar
  \psi}^{\prime}({{\vec r}^{\prime}\over L})|_{{\vec r}^{\prime}={\vec
  r}}
\end{equation}
where the gradient and Laplacian are with respect to ${{\vec
    r}^{\prime}\over L}$

Since ${\bar A}(|{\vec r} - {\vec r}^{\prime}|)\rightarrow 0$ as
$|{\vec r} - {\vec r}^{\prime}|\rightarrow\infty$, ${|{\vec r} - {\vec
    r}^{\prime}|\over L}<< 1$ for large L and we can truncate the
series in eq.(\ref{expan}) after terms of the second order. Inserting
eq.(\ref{drop}) and eq.(\ref{expan}) into eq.(\ref{eulern}) we obtain
\begin{eqnarray}
\label{eulern2}
& R^{d}\bigl\lbrack \sum_{n} \exp(i{\vec k}_{o,n}\cdot{\vec
  r}){\hat{\bar A}}(|{\vec k}_{o}|){\bar \psi}^{\prime}({{\vec r}\over
  L}) - \nonumber \\ 
&G(h,T)\sum_{n}\exp(i{\vec k}_{o,n}\cdot{\vec r}){1\over
  L^{2}}\nabla^{2}{\bar \psi}^{\prime}({{\vec r}\over L}) +
  b(h,T)(\sum_{n}\exp(i{\vec k}_{o,n}\cdot{\vec r}))^{2}{\bar
  \psi}^{\prime 2}({{\vec r}\over L})\bigr\rbrack = 0
\end{eqnarray}
where the Laplacian is with respect to ${{\vec r}\over L}$ and
\begin{equation}
\label{g}
G(h,T) = \int |{\vec r} - {\vec r}^{\prime}|^{2}{\bar A}(|{\vec r} -
{\vec r}^{\prime}|)d({\vec r} - {\vec r}^{\prime}) < 0.
\end{equation}
The term involving the gradient is zero as it reduces to a
function proportional to $\nabla_{\vec k}{\bar{\hat A}}(|{\vec
  k}|)|_{{\vec k}={\vec k}_{o}}$ and $|{\vec k}_{o}|$ is the global
minimum of ${\hat{\bar A}}(|{\vec k}|)$. Since we are near a spinodal 
${\hat{\bar A}}(|{\vec k}_{o}|) = \epsilon<<1$. We can assume\cite{KL} 
that ${\bar \psi}^{\prime}({{\vec r}\over L})\sim \epsilon$ as long as 
$L\sim \epsilon^{-1/2}$ Hence 
\begin{equation}
\label{spsi}
{\bar \psi}^{\prime}({{\vec r}\over L}) = \epsilon \Psi({{\vec x}\over
  R\epsilon^{-1/2}})
\end{equation}
where the scaling of ${\vec x}$ with respect to both $\epsilon^{-1/2}$
and $R$ is explicit. For further discussion of this scaling see section
V. The argument of AM is now invoked to limit the
${\vec k}_{o,n}$ to lie on equilateral triangles. The critical or
nucleating droplet then has the following form: In the interior it is
periodic with a triangular or hexagonal structure in two dimensions
and a bcc or layered hexagonal or triangular plane structure in three
dimensions. These structures are modulated by an envelope which
satisfies the equation
\begin{equation}
\label{eulern3}
G(h,T)\nabla^{2}{\bar \psi}^{\prime}(\epsilon^{1/2}{\vec r}) +
 {\hat{\bar A}}(|{\vec k}_{o}|){\bar \psi}^{\prime}(\epsilon^{1/2}{\vec r}) +
 b(h,T){\bar \psi}^{\prime 2}(\epsilon^{1/2}{\vec r}) = 0
\end{equation} 
where the Laplacian is with respect to ${\vec r}$. The boundary
conditions are that ${\bar \psi}^{\prime}(\epsilon^{1/2}{\vec
  r})\rightarrow 0$ when $|{\vec r}|\rightarrow \infty$ and that 
${d\over d|{\vec r|}}{\bar \psi}^{\prime}(\epsilon^{1/2}{\vec r})=0$
at $|{\vec r}|=0$ The first condition is simply a statement that the
droplet is localized. The second boundary condition assures that the
droplet has no unphysical kinks at its center.\cite{lang} 
Note also that the
structure of eq.(\ref{eulern3}) implies that the solution is a function
of $|{\vec r}|$. Since the sign of $b(h,T)$ merely sets the sign of 
${\bar \psi}^{\prime}(\epsilon^{1/2}|{\vec r}|)$ we lose no generality
by assuming that $b(h,T)<0$.

From the discussion in Section III this droplet will be unstable, not
just on its surface as in classical nucleation\cite{lang}, but in its
interior. We can see this explicitly by finding the eigenvector
associated with the negative eigenvalue\cite{lang} of the operator
obtained by a linear stability analysis about the critical
droplet. That is we want the solution of 
\begin{equation}
\label{eulern4}  
\int {\bar A}(|{\vec r} - {\vec r}^{\prime}|)w({\vec r})^{\prime}d{\vec
  r}^{\prime} - 2|b(h,T)|{\bar \psi}({\epsilon}^{1/2}{\vec
  r})w({\vec r}) = \lambda w({\vec r})
\end{equation}
where ${\bar \psi}(\epsilon^{1/2}{\vec r})$ is the solution to
eq.(\ref{eulern}) given in eq.(\ref{drop}).

We assume that the eigenvector has the form
\begin{equation}
\label{ev}
w({\vec r}) = \sum_{n}\exp(i{\vec k}_{o}\cdot {\vec
  r})W(\epsilon^{1/2}{\vec r})
\end{equation}
Employing the same arguments we used to obtain eq.(\ref{eulern3}) we
find that the eigenvector has the same interior structure as the
critical droplet and an envelope that is the solution of
\begin{equation}
\label{evenvel}
G(h,T)\nabla^{2}W(\epsilon^{1/2}{\vec r}) + \epsilon W(\epsilon^{1/2}{\vec
 r}) - 2|b(h,T)|{\bar \psi}^{\prime}(\epsilon^{1/2}{\vec
 r})W(\epsilon^{1/2}{\vec r}) = \lambda W(\epsilon^{1/2}{\vec r})
\end{equation}
where we have set ${\hat{\bar A}}(|{\vec k}_{o}|) = \epsilon$.

In $d=1$ the solution of eq.(\ref{eulern3}) can easily be seen by
substitution to be
\begin{equation}
\label{1d}
{\bar \psi}^{\prime}(\epsilon^{1/2}x) = {D\epsilon^{1/2}
   \over \cosh^{2}(\alpha \epsilon^{1/2} x)}
\end{equation}
where $\alpha$ and $D$ are constants. In $d=3$ eq.(\ref{eulern3}) has
been solved numerically\cite{uk1}. The solution is radially symmetric,
has its maximum at the origin and decreases to zero as $|{\vec
  r}|\rightarrow \infty$ Hence eq.(\ref{evenvel}) has the form of a
Schr\"odinger equation with a potential $V(|{\vec r}|)$ given by
\begin{equation}
\label{pot}
V(|{\vec r}|) = {\hat{\bar A}}(|{\vec k}_{o}|) - 2{\bar
    \psi}^{\prime}(\epsilon^{1/2}|{\vec r}|)
\end{equation}

From the discussion above, this potential is a shallow well. The
lowest eigenvalue will correspond to a bound state and hence will be
negative. Moreover the eigenvector corresponding to the bound state
will be centered in the center of the well. From the work of
Langer\cite{lang} we know that the eigenstate that corresponds to the
negative eigenvalue is an unstable mode of the droplet. The fact that
the bound state is centered in the center of the well is confirmation
that the entire droplet is unstable consistent with the adaptation of
the AM argument. This form of the eigenvector was also seen
numerically in Ising models.\cite{uk1}.

The dominant contribution to the probability of the occurance of a
critical droplet comes from inserting the solution to the
Euler-Lagrange equation (eq.(\ref{eulern2})) into the expression for the
partition function. The details of this caluclation for liquid-gas  
and magnetic systems can be found in Langers paper for classical
nucleation\cite{lang} and in Unger and Klein for spinodal
nucleation.\cite{UK} The adaptation to liquid-solid nucleation near
the spinodal of the supercooled liquid of the saddle point part of the
calculation is straightforward.\cite{KL} The free energy barrier to
nucleation then is given by 
\begin{equation}
\label{fe}
F(h,T) = R^{d}\int{\bar A}(|{\vec r} - {\vec r}^{\prime}|){\bar
  \psi}(\epsilon^{1/2}{\vec r}){\bar \psi}(\epsilon^{1/2}{\vec
  r}^{\prime})d{\vec r}d{\vec r}^{\prime} - |b(h,T)|\int {\bar
  \psi}^{3}(\epsilon^{1/2}{\vec r})d{\vec r}
\end{equation}
where ${\bar \psi}(\epsilon^{1/2}{\vec r})$ is the solution to
eq.(\ref{eulern}) From eqs. \ref{drop} and \ref{spsi} we have   
\begin{eqnarray}
\label{fe1}
F(h,T) =&R^{d}\epsilon^{3}\int{\bar A}^{\prime}(|{\vec r} - {\vec
  r}^{\prime}|)\bigl\lbrack \sum_{n}\exp(i{\vec
  k}^{\prime}_{o,n}\cdot{\vec r})\Psi(\epsilon^{1/2}{\vec
  r}))\bigr\rbrack \bigl\lbrack \sum_{n}\exp(i{\vec
  k}^{\prime}_{o,n}\cdot{\vec r}^{\prime})\Psi(\epsilon^{1/2}{\vec
  r}^{\prime})\bigr\rbrack d{\vec r}d{\vec r}^{\prime}\nonumber\\ 
   &-|b(h,T)|\int\bigl\lbrack \sum_{n}\exp(i{\vec k}^{\prime}_{o,n}\cdot{\vec
  r})\bigr\rbrack^{3}\Psi^{3}(\epsilon^{1/2}{\vec r})d{\vec r}
\end{eqnarray} 
where ${\bar A}(|{\vec r} - {\vec r}^{\prime}|) = \epsilon{\bar
  A}^{\prime}(|{\vec r} - {\vec r}^{\prime}|)$ and ${\vec
  k}^{\prime}_{o,n} = R{\vec k}_{o,n}$

When $|{\vec k}_{o,n}|\neq 0$
\begin{equation}
\label{supp}
\int \exp(i{\vec k}_{o,n}\cdot{\vec r})\Psi(\epsilon^{1/2}{\vec
  r})d{\vec r}\sim {1\over \xi^{d}} = \epsilon^{d/2}
\end{equation}
where $\xi = \epsilon^{-1/2}$ is 
the correlation length near the spinodal in scaled units. 
Therefore, the dominant contribution to the cubic term in
  eq.(\ref{fe1}) will be given by those terms in the sum
  $(\sum_{n}e^{i{\vec k}_{o,n}\cdot{\vec r}})^{3}$ which are spatial
  constants. The same analysis can be done for the quadratic term in
  eq.(\ref{fe1}). The dominant contribution will come from the ${\vec
    k}_{o,n}$ vectors that appear with opposite signs in the sums and
  result in exponentials of the form $\exp({\vec k}_{o,n}\cdot({\vec
    r}-{\vec r}^{\prime}))$. These terms will be of order $\xi^{d}$ and
  all other terms will be reduced by a factor of $1/\xi^{d}$ as in
  eq.(\ref{supp}).  

The free energy barrier is then 
\begin{equation}
\label{feb}
F(h,T) = MR^{d}\epsilon^{3-d/2}
\end{equation}
where $M$ is a constant. The probability of a critical droplet is 
\begin{equation}
\label{pcd}
P_{D} = g(h,T)\exp(-MR^{d}\epsilon^{3-d/2})
\end{equation}
where $g(h,T)$ is small compared to the exponential.\cite{lang,UK}

The exponential dominates the probability of a nucleation or critical
droplet. To get a feel for the magnitude of the argument of the
exponential we turn to consideration of the so-called Ginsburg
criterion for the validity of a mean-field treatment.\cite{bin} In MF
systems fluctuations can be ignored when calculating thermodynamic
quantities.\cite{bin} Ginsburg pointed out that this implies that the
fluctuations of the order parameter must be small compared to its mean
value. That is
\begin{equation}
\label{gin}
{\xi^{d}\chi_{T}\over \xi^{2d}\phi^{2}} << 1
\end{equation}
where $\xi$ is the correlation length as above, $\chi_{T}$ is the
susceptibility and $\phi$ is the order parameter. Using the MF
exponents for the spinodal, $\chi_{T}\sim
\epsilon^{-1}$, $\phi\sim \epsilon$ and $\xi\sim R\epsilon^{-1/2}$,
where we have made the $R$ dependence of the correlation length
explicit, we have
\begin{equation}
\label{gin1}
{\epsilon^{-1}\over R^{d}\epsilon^{2-d/2}} << 1
\end{equation}
or
\begin{equation}
\label{gin2}
R^{d}\epsilon^{3-d/2} >> 1
\end{equation}
for MF theory to be a good approximation. If the left hand side of
eq.(\ref{gin2}) is infinite then MF theory is exact. Since spinodals
are MF objects and pseudo-spinodals affect the physics only if the
system is NMF we require the condition in eq.(\ref{gin2}) to hold if
we are to see the spinodal nucleation process described above.  

If $R >> 1$ then $\epsilon$ can be small and the spinodal can be
approached ($\epsilon << 1$) with the Ginsburg criterion,
$R^{d}\epsilon^{3-d/2}$, still valid. For Ising models the spinodal
can be seen by measurements of the isothermal susceptibility which
will diverge as $R\rightarrow \infty$ and $\epsilon\rightarrow
0$.\cite{heerm} In the supercooled liquid the static structure factor
is known rigorously to diverge at $|{\vec k}|\neq 0$.\cite{grewe} However,
in this case the situation is more complicated and the 
measured divergence is suppressed for $d >1$.\cite{klet}

The implications of these investigations is that the argument of AM,
while it does not predict the structure of stable crystals, can be
adapted to determine the structure of at least one kind of nucleation
droplet near the pseudo-spinodal in liquids with long range
interactions. Questions have been raised in the work of GM
about the possibility of other forms of nucleation droplets near the
spinodal. In the next section we will address this question. 

\section{Uniqueness of critical droplet}

The question of the uniqueness of a critical droplet, given the
thermodynamic parameters that specify the metastable state, has not
been fully resolved. This is true even in phase transitions , such as
gas-liquid, where there is no spatial symmetry breaking. The
resolution of this problem is even more difficult when, as in the
nucleation of the crystal from the liquid, the spatial symmetry
changes. To completely answer this question for a given $T$ and $h$,
all solutions of the Euler-Lagrange equation resulting from setting
the functional derivative of $H(\rho)$ found in eq.(\ref{ham2}) would
have to be found and the free energy cost of each solution
evaluated. This is a formidable task that has yet to be done either
analytically or numerically. 

In the simpler liquid-gas nucleation process, although it has not been
proven, the idea that there is only one saddle point separating the
metastable and stable state seems quite reasonable. In the
liquid-solid transition multiple saddle points with different spatial
symmetries seems more plausible. In this regard, Groh and
Mulder(GM)\cite{GM} have performed a bifucation analysis of
supercooled liquids near the spinodal. They found that the first order
bifucation analysis yields only a bcc solution but that the second
order analysis results in a solution with an fcc symmetry. They also
found that within the  mean-field Landau theory the bcc structure has a
lower maximum of the free energy than the fcc structure. The word
structure was used rather than phase since, as we have seen in Section
III, these structures are unstable. 
Groh and Mulder however, do not consider nucleation. It is the purpose
of this section to provide an argument, unfortunately not a proof,
that as the spinodal is
approached the bcc nucleation droplet found in the previous section
dominates the nucleation process. 

We begin by noting that it is
necessary to have the linear size of the critical or
nucleation droplet equal to or larger than the correlation
length. Particles separated by a scale smaller than the correlation
length are highly correlated. Therefore, the statistical fluctuations
that begin the evolution up a saddle point hill should involve regions
that are at least the correlation length size size. The converse is
not true. Critical droplets can be large compared to the correlation
length since critical droplets are rare compared to the fluctuations
that set the scale for correlations. It has been the case in all
simulations that have looked at nucleation that the critical droplet
size is either equal to or larger than the correlation
length.\cite{hcks,gunt} This being the case then the critical droplet
profile $\psi({\vec x})$ can be approximated by the form     
\begin{equation}
\label{drpt}
\psi({\vec x}) = \sum_{n}\exp(i{\vec k}_{n}\cdot{\vec r})\Phi(|{{\vec
  r}\over L}|)
\end{equation}
where the ${\vec k}_{n}$ are the entire set of reciprocal lattice vectors
and $L\geq \xi$. This form of the droplet clearly is not exact. We
would expect some effect of the fact that the envelope will decay to
zero as $|{{\vec r}\over L}|\rightarrow \infty$ on the
symmetry. However, that will be in the tail of the droplet, which is
expected to have exponential decay\cite{UK,KL,gunt} 
(also see eq.(\ref{1d}) and will have
negligible effect on the interior symmetry. Near the spinodal where
the correlation length $\xi$ diverges, the interior of the droplet will
be unaffected by the envelope. In addition any scaling should also be
unaffected by the approximation made at the edge of the droplet. 

With arguments essentially identical to those of Section IV the
envelope obeys an equation similar to eq.(\ref{eulern2}). 

\begin{eqnarray}
\label{drpteq}
&\sum_{n}\exp(i{\vec k}_{n}\cdot{\vec r}){\hat{\bar A}}(|{\vec
    k}_{n}|)\Phi({{\vec r}\over L})
-|G(h,T)|\sum_{n}\exp(i{\vec k}_{n}\cdot{\vec r})\nabla^{2}\Phi({{\vec
    r}\over L}) +\nonumber \\  
&b(h,T)(\sum_{n}\exp(i{\vec k}_{n}\cdot{\vec r}))^{2}\Phi^{2}({{\vec
    r}\over L}) + c(h,T)(\sum_{n}\exp(i{\vec k}_{n}\cdot{\vec
    r}))^{3}\Phi^{3}({{\vec r}\over L}) = 0 
\end{eqnarray}
where the Laplacian is with respect to ${\vec r}$ and $G(h,T)$,
$b(h,T)$ and $c(h,T)$ are independent of ${\vec r}$ and ${\vec k}$.
The Laplacian term is of order ${1\over L^{2}}$ and $L\geq \xi\sim
\epsilon^{-1/2}$. We have restricted our considerations to
Hamiltonians that include terms up to $\Phi^{4}({{\vec r}\over L})$
neglecting higher orders in $\Phi$. These higher order terms can be
included with no essential change in the argument.  If the term 
\begin{equation}
\label{term1}
\sum_{n}\exp(i{\vec k}_{n}\cdot{\vec r}){\hat{\bar A}}(|{\vec k}_{n}|)
\rightarrow 0
\end{equation}
slower than ${1\over L^{2}}$ as the spinodal is approached, 
then the Laplacian term can be ignored. 
This would result, assuming that symmetry constraints could be
satisfied, in $\Phi({{\vec r}\over L})$ being a spatial
constant. Since nucleation droplets are, almost by definition,
localized this result would clearly be unsatisfactory. The only way
that the term in eq.(\ref{term1}) can go to zero at all is if the sum
is limited to terms in which $|{\vec k}_{n}| = |{\vec k}_{o}|$ where
${\vec k}_{o}$ has the same meaning as in previous sections. (See
eq.(\ref{eulern2})) 

Therefore, we must have
\begin{equation}
\label{term1a}
\sum_{n}\exp(i{\vec k}_{n}\cdot{\vec r}){\hat{\bar A}}(|{\vec k}_{n}|)
= \sum_{n}\exp(i{\vec k}_{o,n}\cdot {\vec r}){\hat{\bar A}}(|{\vec
  k}_{o,n}|)\sim \epsilon
\end{equation}
as this is the only possible way that this term can go to zero as
$\epsilon\rightarrow 0$. Since the terms linear in $\Phi$ must scale
the same way this implies that ${1\over L^{2}}\sim \epsilon$ so that
$L\sim \epsilon^{-{1\over 2}}$ or $L\sim \xi$. If b(h,T) is not zero
in eq.(\ref{drpteq}) then $\Phi({{\vec r}\over L})$ must scale as
$\epsilon$. If it were to scale as $\epsilon^{x}$ with $x>1$ then the
non-linear terms in  $\Phi$ in eq.(\ref{drpteq}) can be ignored
relative to the linear terms. This results in a solution of the form
\begin{equation}
\label{soln5}
\Phi({{\vec r}\over L}) = \Phi(\epsilon^{1/2}{\vec r}) =
C_{1}\exp(\epsilon^{1/2}{\hat n}\cdot {\vec r}) +
C_{2}\exp(-\epsilon^{1/2}{\hat n}\cdot {\vec r})
\end{equation} 
where ${\hat n}$ is a unit vector whose direction is arbitrary.
For the droplet to be localized the constant $C_{1}$ must be set to
zero. This results in a violation of the boundary condition that 
the derivative with respect to $|{\vec r}|$ is zero at $|{\vec r}| =0$.
Hence this droplet is physically ruled out. If $x<1$ then the
linear terms in $\Phi$, including the Laplacian term, can be ignored  
resulting again in a non-local solution. In general nucleation from
the liquid to the solid is described by Hamilitonians in which
$b(h,T)\neq 0$ and hence the AM argument can be used to fix the
spatial symmetry in the interior of the critical droplet. 
Note that $b(h,T)$ is a function of $h$ and $T$ and is therefore
specified only by the parameters that define the metastable state. 

Suppose that $b(h,T) = 0$. In this case, arguments similar to the ones
above imply that $\Phi({{\vec r}\over L})\sim \epsilon^{1/2}$. The
$\Phi^{4}$ term in the Hamiltonian is now relevant and the 
nucleation barrier will now scale as $R^{d}\epsilon^{2-d/2}$. 
In addition a simple extension of AM indicates that the droplet
symmetry will now be a square lattice in $d=2$ and an fcc lattice, or
stacked layers of square lattices, in $d=3$. 

It is interesting to ask why, except for a particular symmetry of the
Hamiltonian, the critical droplet has a bcc symmetry in $d=3$ when the
bifurcation analysis of GM indicates that an fcc bifurcation is also
allowed even when the coefficient ($b(h,T)$) of the cubic term in the
Hamiltonian is nonzero. Repeating the GM analysis is beyond the scope
of this paper however, the key to their construction of the fcc
bifurcation is the use of lattice vectors for which $|{\vec k}|\neq
|{\vec k}_{o}|$. The argument we used made use of 1) the requirement
that the droplet be localized, 2)the droplet has no kinks at its center 
and 3)the assumption that the linear
droplet size $L$ was greater than or equal to the correlation length
$\xi$. These conditions taken together resulted (see equations
\ref{term1} and \ref{term1a}) in a restriction of the reciprocal
lattice vectors to those for which $|{\vec k}| = |{\vec
  k}_{o}|$. These conditions, while necessary for specifying the
nucleation problem are not needed in the bifurcation analysis of GM. 
Hence, even though the fcc bifurcation is allowed it will
not occur in a critical droplet as the spinodal is approached.

It is important to emphasize that we have restricted our consideration
to Hamiltonians that are in the same class as that considered by
AM. In density functional language this implies that all direct
correlation functions higher than the pair are zero. Whether the
inclusion of higher order correlation functions changes the result
remains to be seen.

\section{Summary and discussion}

We have argued that the prediction of a stable bcc phase near a
spinodal, or pseudo-spinodal, of a supercooled liquid by AM is not
correct. The predicted bcc phase is unstable. However we have used
the AM argument to show that the nucleation or critical droplet near a
spinodal does have a bcc structure. The same argument that was used to
show that the AM ``crystal'' was unstable is also used to show that
the critical droplet is unstable. This is however, physically
reasonable for a critical droplet near a spinodal.\cite{KL,UK} 

It should also be noted that droplets like the ones predicted in this
paper were found in molecular dynamics simulations of supercooled Lennard-Jones  
fluids.\cite{yang} In addition it is known from theoretical studies of
the liquid-gas transition that as one moves away from the
pseudo-spinodal the droplet develops a core that appears almost
classical.\cite{uk1} This phenomenon is consistent with the results of
a density functional calculation of liquid solid nucleation by Shen
and Oxtoby.\cite{so}   

Many systems of technological importance have long range, or effective
long range interactions. These include polymers\cite{bin}, neutral
plasmas and metals.\cite{shenoy} Since critical droplets are at least
the size of the correlation length, and all statistical lengths, such as
the droplet diameter, are measured in units of the interaction range
$R$, classical nucleation, where the surface tension is assumed to
remain non-zero\cite{lang} will be strongly suppressed. In these
systems, spinodal nucleation will dominate the metastable state
decay. Since the structure of the interior of a spinodal nucleation
droplet is not the stable crystal as would be expected in the
classical process the evolution of this droplet as it grows will be an
important step in determining the structure of the stable or
metastable crystal.

\centerline{\bf ACKNOWLEGEMENTS}

It is a pleasure for me to acknowledge interesting and stimulating
discussions with E. Matolla, T. Lookman, A. Saxena, F.J. Alexander and
D. Hatch. This work was performed under the auspicies of the DOE at
Los Alamos National Laboratory under grants DE-FG02-95ER14498 and 
LDRD-DR-2001501. It is also a pleasure for me to acknowledge the
support and hospitality of the Center for Non-Linear Studies at Los
Alamos National Laboratory. This work was supported by the Department
of Energy, under contract W-7405-ENG-36



\end{document}